\documentclass[twocolumn,floatfix]{revtex4}
\pdfoutput=1
\usepackage{graphicx}
\usepackage{dcolumn}
\usepackage{longtable}
\usepackage{amsmath}
\usepackage{amssymb}
\usepackage{bm}
\begin{document} 

\title{Seven years of the proxy-SU(3) shell model symmetry}

\author
{Dennis Bonatsos$^1$, Andriana Martinou$^1$, S.K. Peroulis$^1$, T.J. Mertzimekis$^2$, and N. Minkov$^3$ }

\affiliation
{$^1$Institute of Nuclear and Particle Physics, National Centre for Scientific Research ``Demokritos'', GR-15310 Aghia Paraskevi, Attiki, Greece}

\affiliation
{$^2$  Department of Physics, National and Kapodistrian University of Athens, Zografou Campus, GR-15784 Athens, Greece}

\affiliation
{$^3$Institute of Nuclear Research and Nuclear Energy, Bulgarian Academy of Sciences, 72 Tzarigrad Road, 1784 Sofia, Bulgaria}

\begin{abstract}
The proxy-SU(3) symmetry was first presented in HINPw4 in Ioannina in May2017, justified within the Nilsson model and applied to parameter-free predictions of the collective variables $\beta$  and $\gamma$ in medium-mass and heavy nuclei. Major steps forward, including the connection of the proxy-SU(3) symmetry to the shell model, the justification  of the dominance of highest weight states  in terms of the short range nature of the nucleon-nucleon interaction, as well as the first proposal of appearance of islands of shape coexistence on the nuclear chart, have been presented in HINPw6 in Athens in May 2021. The recently hot topic of the prevalence of triaxial shapes in heavy nuclei will also be briefly outlined in the proxy-SU(3) framework.
  
 \end{abstract}

\maketitle
\section{Introduction}

SU(3) symmetry has been widely used in nuclear physics \cite{Kota2020}. In 2017, a new approximate manifestation of the SU(3) symmetry in the nuclear shell model, called the proxy-SU(3) symmetry, has been introduced \cite{Bonatsos2017a,Bonatsos2017b,Bonatsos2023a} and presented at the HINPw4 workshop in Ioannina \cite{HINP2017Bon,HINP2017Assim,HINP2017Mart,HINP2017Sar}. Further developments within the framework of the proxy-SU(3) symmetry have been reported in 2021 at the HINPw6 workshop in Athens \cite{HINP2021SM,HINP2021SC,HINP2021hw}. In the present paper we briefly review the foundations of the proxy-SU(3) symmetry in sections \ref{prox} and
\ref{hwirreps}, while its major achievements are briefly described in sections \ref{bg} and \ref{isc}, with a new possible application outlined in section \ref{triaxial}.  

\section{The 0[110] proxies} \label{prox}

SU(3) symmetry has been present in the nuclear spherical shell model \cite{Heyde1990,Talmi1993} since its introduction in 1949 \cite{Mayer1948,Mayer1949,Haxel1949,Mayer1955}, which was based on the 3-dimensional isotropic harmonic oscillator (3D-HO), the shells of which, characterized by the number of excitation quanta $n$, are known to bear the unitary symmetries U((n+1)(n+2)/2) \cite{Wybourne1974,Moshinsky1996,Iachello2006}, which possess SU(3) subalgebras \cite{Bonatsos1986}. However, the SU(3) symmetry is broken beyond the $n=2$ ($sd$) shell, because of the spin-orbit interaction, which is necessary in order to explain the experimentally observed magic numbers 2, 8, 20, 28, 50, 82, 126, \dots, at which nuclei appear to be extremely stable. Within each shell, the spin-orbit interaction pushes the orbitals bearing the highest eigenvalue of the total angular momentum $j$ to the shell below. Thus each shell remains with its initial orbitals, minus the ones deserting it towards the shell below, plus the orbitals invading from the shell above, called the intruder orbitals. In the case of the $sdg$ shell, for example, shown in Fig. 1 of Ref. \cite{Bonatsos2017a}, the $1g_{9/2}$ orbital leaves for the shell below, while the $1h_{11/2}$ is invading from above, forming with the remnants of the $sdg$ shell (the $1g_{7/2}$, $2d_{5/2}$, $2d_{3/2}$, $3s_{1/2}$ orbitals) the nuclear 50-82 shell.   

Do the deserting orbitals $1g_{9/2}$ and the invading orbitals $1h_{11/2}$ bear any similarity? Indeed they do, but in order to unmask it one has to use the deformed basis introduced by Nilsson \cite{Nilsson1955,Nilsson1995} in 1955, in order to be able to accommodate deformation within the nuclear shell model. Indeed, soon after the introduction of the nuclear shell model it was realized that it could not explain the large nuclear quadrupole moments observed away from closed shells. Rainwater \cite{Rainwater1950} in 1950 realized that spheroidal nuclear shapes are energetically favored over spherical ones, and soon thereafter, in 1952, Bohr and Mottelson \cite{Bohr1952,Bohr1953} introduced the nuclear collective model \cite{Bohr1998a,Bohr1998b}, in which nuclear properties are described in terms of the collective variables $\beta$ and $\gamma$, corresponding to the departure from sphericity and from axial symmetry respectively. These developments led Nilsson in 1955 to introduce a deformed shell model with axial symmetry \cite{Nilsson1955,Nilsson1995}, in which the single particle levels are described by the quantum numbers $\Omega [N n_z \Lambda]$, where $N$ is the number of oscillator quanta, $n_z$ is the number of quanta along the $z$-axis of axial symmetry, and $\Lambda$ ($\Omega$) is the projection of the orbital (total) angular momentum on the $z$-axis. Looking into the details (see Table 4 of Ref. \cite{Bonatsos2020b}), one can easily see that the levels forming the $1g_{9/2}$ orbital are in one-to-one correspondence with the levels forming the $1h_{11/2}$ orbital, bearing the same projections of orbital  and total angular momentum and differing only by one quantum in the $z$-direction. Thus there pairs of levels differ by $\Delta \Omega [\Delta N \Delta n_z \Delta\Lambda]=0[110]$, and are called 0[110] pairs. Only the $1h_{11/2}$ level with the highest projection of the total angular momentum has no counterpart among the $1g_{9/2}$ levels, as one can see in Table 4 of Ref. \cite{Bonatsos2020b}. 

Empirical considerations of data for double differences of binding energies have shown \cite{Cakirli2005,Cakirli2010} that proton-neutron 0[110] pairs correspond to maximization of the proton-neutron interaction, a result corroborated by relativistic mean field calculations \cite{Stoitsov2007}. It has also been proved that 0[110] pairs possess the highest orbital overlaps \cite{Bonatsos2013}. In addition to corroborating the maximization of the proton-neutron interaction, maximal overlaps also indicate the great similarity between the two orbitals. It becomes therefore plausible to try to replace the intruder $1h_{11/2}$ levels by their $1g_{9/2}$ 0[110] counterparts, which can then serve as their ``proxies''. 

The accuracy of this replacement has been checked in detail \cite{Bonatsos2017a} within the Nilsson model, as well as within the spherical shell model \cite{Martinou2020,Bonatsos2020b}. Numerical studies have shown that Nilsson levels are only slightly influenced by this replacement \cite{Bonatsos2017a}. Considerations within the spherical shell model have shown that the two sets of levels, shown in Fig. 1 of Ref. \cite{Martinou2020}, are connected by a simple unitary transformation \cite{Martinou2020}. The $1h_{11/2}$ level with the highest projection of the total angular momentum, which  has no counterpart among the $1g_{9/2}$ levels, is seen in the standard Nilsson diagrams \cite{Lederer1978} to lie at the top of the relevant shell, thus it is empty for most nuclei using this shell, and therefore its influence is negligible. 

\section{The highest weight irreducible representations} \label{hwirreps}

A major step forward in nuclear structure theory has been taken in 1958 by Elliott \cite{Elliott1958a,Elliott1958b,Elliott1963}, who classified the states in the $sd$ nuclear shell, bearing the U(6) overall symmetry, by  using its noncanonical chain of subalgebras SU(3)$\supset$SO(3). The irreducible representations of SU(3) have been labeled by Elliott as $(\lambda,\mu)$, with $\lambda=f_1-f_2$ and $\mu=f_2$, where $f_1$, $f_2$ are the number of boxes in the first and second line of the relevant Young diagram respectively. The irreps of SO(3) are labeled by $L$, the orbital angular momentum. Multiple states with the same $L$ occurring within the same SU(3) irrep $(\lambda,\mu)$ are distinguished by the ``missing'' quantum number $K$ in the noncanonical chain SU(3)$\supset$SO(3), called the Elliott quantum number in the literature. In this way, nuclear states within each SU(3) irrep 
$(\lambda,\mu)$ are distributed into bands, characterized by different $K$, each band containing a series of levels with increasing $L$. 

A question which immediately comes is the one of choosing the SU(3) irrep lying lowest in energy, to which the ground state band will belong. For decades it has been widely believed that the lowest lying irrep is the one bearing the highest eigenvalue \cite{Iachello2006} of the second order Casimir operator of SU(3), 
\begin{equation}
C_2(\lambda,\mu)={2\over 3} (\lambda^2 +\mu^2+\lambda \mu + 3\lambda +3 \mu), 
\end{equation}
which is connected to the quadrupole-quadrupole interaction through the relation \cite{Elliott1958a,Elliott1958b,Elliott1963,Draayer1993}
\begin{equation}
\hat Q\cdot \hat Q = 4 \hat C_2 -3 \hat L^2. 
\end{equation}
It was therefore believed that the highest $C_2$ eigenvalue corresponds to the strongest quadrupole-quadrupole interaction and therefore to the  lowest energy. 

\begin{table}[htb]\label{Tab6}

\caption{Highest weight SU(3) irreps (labeled by hw) and highest $C_2$ irreps (labeled by C) for U(n), n=6, 10, 15. Highest weight (hw) irreps differing from their highest $C_2$ counterparts are shown in boldface.  
Adapted from Ref. \cite{SDANCA2017Sar}, where results for n=21, 28, 36 can also be found. See section \ref{hwirreps} for further discussion.}

\bigskip
\begin{tabular}{ r l r r r r r r  } 

\hline

\hline
   &             & 8-20 & 8-20 & 28-50 & 28-50& 50-82 & 50-82 \\
   &             & sd   &  sd  & pf    &  pf  & sdg   &  sdg  \\
M  & irrep       & U(6) & U(6) & U(10) & U(10)& U(15) & U(15) \\
   &             & hw   & C    & hw    &  C   & hw    &  C    \\
 0 &             &(0,0) &(0,0) &(0,0)  &(0,0) &(0,0)  &(0,0)  \\  
 1 & [1]         &(2,0) &(2,0) & (3,0) &(3,0) & (4,0) &(4,0)  \\
 2 & [2]         &(4,0) &(4,0) & (6,0) &(6,0) & (8,0) &(8,0)  \\
 3 & [21]        &(4,1) &(4,1) & (7,1) &(7,1) &(10,1) &(10,1) \\
 4 & [$2^2$]     &(4,2) &(4,2) & (8,2) &(8,2) &(12,2) &(12,2) \\
 5 & [$2^2$1]    &(5,1) &(5,1) &(10,1) &(10,1)&(15,1) &(15,1) \\
 6 & [$2^3$]     &(6,0) & (0,6)&(12,0) &(12,0)&(18,0) &(18,0) \\
 7 & [$2^3$1]  &{\bf(4,2)}&(1,5)&(11,2)&(11,2)&(18,2) &(18,2) \\
 8 & [$2^4$]     &(2,4) & (2,4)&(10,4) &(10,4)&(18,4) &(18,4) \\
 9 & [$2^4$1]    &(1,4) & (1,4)&(10,4) &(10,4)&(19,4) &(19,4) \\
10 & [$2^5$]     &(0,4) & (0,4)&(10,4) &(4,10)&(20,4) &(20,4) \\
11 & [$2^5$1]   &(0,2)&(0,2)&{\bf(11,2)}&(4,10)&(22,2)&(22,2) \\
12 & [$2^6$]    &(0,0)&(0,0)&{\bf(12,0)}&(4,10)&(24,0)&(24,0) \\
13 & [$2^6$1]   &     &     &{\bf(9,3)} &(2,11)&(22,3)&(22,3) \\
14 & [$2^7$]    &     &     &{\bf(6,6)} &(0,12)&(20,6)&(20,6) \\
15 & [$2^7$1]   &     &     &{\bf(4,7)} &(1,10)&(19,7)& (7,19)\\
16 & [$2^8$]    &  &      & (2,8) & (2,8)&{\bf(18,8)} & (6,20)\\
17 & [$2^8$1]   &  &      & (1,7) & (1,7)&{\bf(18,7)} & (3,22)\\
18 & [$2^9$]    &  &      & (0,6) & (0,6)&{\bf(18,6)} & (0,24)\\
19 & [$2^9$1]   &  &      & (0,3) & (0,3)&{\bf(19,3)} & (2,22)\\
20 & [$2^{10}$] &  &      & (0,0) & (0,0)&{\bf(20,0)} & (4,20)\\
21 & [$2^{10}$1]&  &      &       &      &{\bf(16,4)} & (4,19)\\
22 & [$2^{11}$] &  &      &       & &{\bf(12,8)} & (4,18)  \\
23 & [$2^{11}$1]&  &      &       & &{\bf(9,10)} & (2,18)  \\
24 & [$2^{12}$] &  &      &       & &{\bf(6,12)} & (0,18)   \\
25 & [$2^{12}$1]&  &      &       & &{\bf(4,12)} & (1,15)  \\
26 & [$2^{13}$] &      &      &   &      &(2,12) & (2,12)   \\
27 & [$2^{13}$1]&      &      &   &      &(1,10) & (1,10)   \\
28 & [$2^{14}$] &      &      &   &      & (0.8) & (0,8)    \\
29 & [$2^{14}$1]&      &      &   &      & (0,4) & (0,4)  \\
30 & [$2^{15}$] &      &      &   &      & (0,0) & (0,0)    \\

\hline

\end{tabular}

\end{table} 

However, our studies \cite{Bonatsos2017b,Martinou2021b} demonstrated that this argument is valid only in the lowest half of each shell. The rule which is valid over the whole shell is that the most symmetric irrep, called the highest weight (hw) irrep in group theoretical language, lies lowest in energy, this being a direct consequence of the Pauli principle and the short-range nature of the nucleon-nucleon interaction \cite{Casten2000}. While the hw irrep is identical to the irrep possessing the highest $C_2$ eigenvalue in the lower half of the shell, the two are in general different in the upper half of the shell, as one can see in Table I. The consequences of this difference are vividly demonstrated in Fig. 1 of Ref. \cite{Bonatsos2017b}. While the highest $C_2$ irreps predict for the square root of $C_2$ a curve vs. particle number symmetric around mid-shell, this is not the case any more if the hw irreps are considered. This fact has important physical consequences, as we shall see in the next sections.

A few comments are in place before leaving this section.

The importance of Elliott's work lies in the fact that it provided a microscopic link between the spherical shell model and nuclear deformation, by pointing out the dominant role of the quadrupole-quadrupole interaction in building nuclear deformation \cite{Elliott1958a,Elliott1958b,Elliott1963}. It should be noticed that the classification of nuclear states in terms of the noncanonical chain SU(3)$\supset$SO(3) by Elliott in 1958 came a few years before the use of the canonical chain SU(3)$\supset$SU(2) for the classification of hadrons according to the eightfold way by Ne'eman \cite{Neeman1961} and Gell-Mann \cite{GellMann1961} in 1961.    

It is worth pointing out that hw irreps have been used instead of their highest $C_2$ counterparts in early works by Hecht and collaborators (see, for example, Table 5 of Ref. \cite{RatnaRaju1973}), but somehow this point had been overlooked over several years afterwards. 

It should be noticed that already in his first paper in 1958 \cite{Elliott1958a} Elliott provides an alternative classification of the $sd$ shell in terms of the O(6)$\supset$O(5)$\supset$SO(3) chain of subalgebras. The implications of this classification have not yet been fully exploited. 

\section{Parameter-independent predictions of the collective variables $\beta$ and $\gamma$} \label{bg}

A connection between the Elliott labels $\lambda$, $\mu$ and the collective variables $\beta$ and $\gamma$ of Bohr and Mottelson has been achieved my making a mapping \cite{Castanos1988,Draayer1989} between the invariant quantities of the two theories, resulting in the relations 
\begin{equation}\label{gamm}
\gamma = \arctan\left( {\sqrt{3} (\mu+1) \over 2\lambda+\mu+3} \right), 
\end{equation}
\begin{equation}
\beta^2 \propto C_2(\lambda,\mu), 
\end{equation}
see Eq. (3) of Ref. \cite{Bonatsos2017b} for the details of the proportionality term in the last equation. From Eq. (\ref{gamm}) one sees that irreps with $\lambda > \mu$  correspond to prolate-like shapes with $\gamma < 30^{\rm o}$,  irreps with $\lambda < \mu$  describe oblate-like shapes with $\gamma > 30^{\rm o}$, while irreps with $\lambda=\mu$ represent maximal triaxiality ($\gamma =30^{\rm o}$). These relations provide parameter-independent predictions for the collective variables $\beta$ and $\gamma$ using only the Elliott labels $\lambda$, $\mu$ of the hw irrep, which itself is determined solely by the Pauli principle and the short-range nature of the nucleon-nucleon interaction. 

The predominance of the hw irreps provided an answer to the long-lasting question \cite{Hamamoto2012} of the dominance of prolate over oblate shapes in the ground states of even-even nuclei. It also predicted a prolate to oblate transition in the heavy rare earths around $N=114$ \cite{Bonatsos2017b,Bonatsos2024}, where a pothole appears in the plot of $\beta$ vs. the neutron number for several series of rare earth isotopes,  as one can see in Fig. 4 of Ref. \cite{Bonatsos2017b}. Its consequences on locating regions of triaxial deformation on the nuclear chart will be considered in section \ref{triaxial}.

It should be noticed that the pseudo-SU(3) symmetry \cite{Arima1969,RatnaRaju1973,Draayer1982}, which is an earlier alternative way of restoring the SU(3) symmetry in the nuclear shells beyond the $sd$ one, provides predictions for $\beta$ and $\gamma$ similar \cite{Bonatsos2020a} to those of proxy-SU(3), despite the fact that pseudo-SU(3) treates the intruder orbitals separately from the ones remaining in each shell, for which it restores the SU(3) symmetry through a different (and more involved) unitary transformation \cite{Castanos1992,Castanos1994}, mapping them to the shell lying below the one under consideration.  

\section{Islands of shape coexistence} \label{isc}

Shape coexistence \cite{Heyde1983,Wood1992,Heyde2011,Bonatsos2023} is said to occur in a nucleus if its ground state band is accompanied by another band lying at similar energy but possessing very different structure. For example, one of the bands can be deformed and the other one spherical, or both of them can be deformed, with one of them being prolate and the other one oblate. 

Proxy-SU(3) brought under attention two sets of nuclear shells \cite{Martinou2021}, the set based on the 3D-HO magic numbers 2, 8, 20, 40, 70, 112, 168, \dots, which corresponds to the absence of the spin-orbit interaction, and the set 6, 14, 28, 50, 82, 126, 184, \dots, which corresponds to the strong presence of the spin-orbit interaction. The interplay of the two sets, in the framework of a dual shell mechanism \cite{Martinou2021}, leads to the parameter-independent prediction that shape coexistence can occur only within the intervals 7-8, 17-20, 34-40, 59-70, 96-112, 146-168, which appear as stripes over the nuclear chart in Fig. 1. A detailed consideration \cite{Bonatsos2023} of theoretical predictions and experimental data led to a collection of nuclei exhbiting shape coexistence, also depicted in Fig. 1. It is seen that these nuclei fall within the stripes predicted by the dual shell mechanism \cite{Martinou2021} developed in the framework of the proxy-SU(3) symmetry. These predictions have been corroborated by covariant density funcional theory calculations \cite{Bonatsos2022a,Bonatsos2022b} and additional empirical considerations \cite{Martinou2023,Bonatsos2023b}.  


\begin{figure*} [htb]

 \includegraphics[width=175mm]{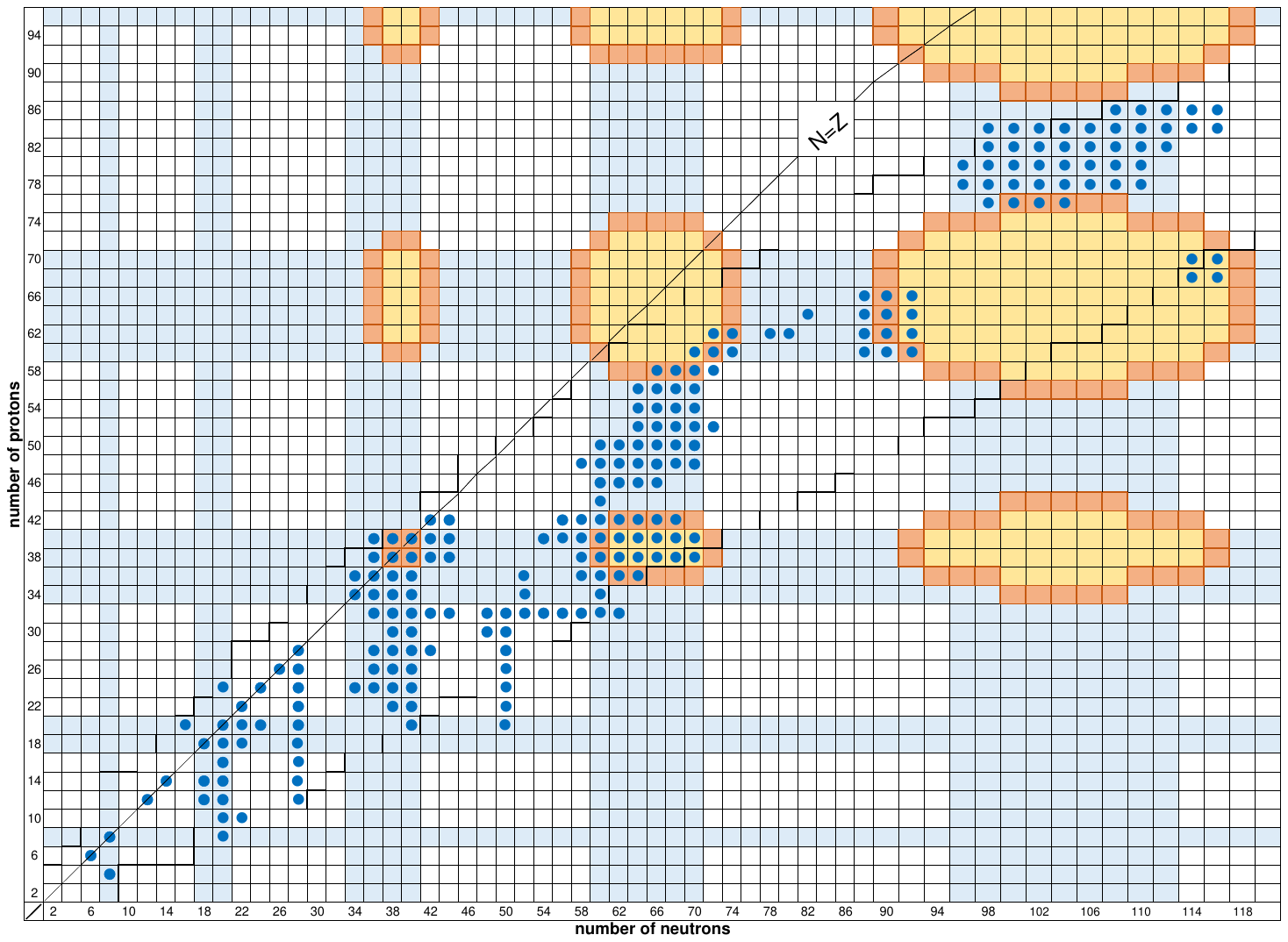}
    \caption{Nuclei exhibiting shape coexistence according to the evidence discussed in Ref. \cite{Bonatsos2023}. Most nuclei fall within the stripes predicted by the dual shell mechanism \cite{Martinou2021} developed in the framework of the proxy-SU(3) symmetry \cite{Bonatsos2017a,Bonatsos2023a}, shown in azure. Contours corresponding to $P\approx 5$ \cite{Casten1987}  are also shown in orange, with deformed nuclei lying within them. Adapted from Ref. \cite{Bonatsos2023}.  See Section \ref{isc} for further discussion.} 
    \label{FR42YY}
\end{figure*}

\section{Regions in which triaxiality should be favored} \label{triaxial}

Recent investigations have raised the question of the appearance of some degree of triaxiality over extended regions of the nuclear chart. Consistent modeling of several observables has shown \cite{Grosse2022} that the variance of the triaxiality variable $\gamma$ is centered around $8^{\rm o}$, in qualitative agreement with the left panel of Fig. 5 of Ref. \cite{Bonatsos2017b}, in which empirical values of $\gamma$ extracted from the data through the Davydov model \cite{Davydov1958,Davydov1959} are shown and compared to the parameter-independent predictions of proxy-SU(3) appearing in the right panel.  Extended calculations within the framework of the triaxial projected shell model (TPSM) \cite{Rouoof2024} have identified 30 nuclei with large triaxiality. The role played by the tensor force in giving rise to triaxiality has been investigated within the Monte Carlo shell model (MCSM) \cite{Tsunoda2021,Otsuka2023}.  

On the other hand, triaxiality has been described within the SU(3)$^*$ symmetry \cite{Dieperink1982,Walet1987} of the interacting boson model-2 (IBM-2) \cite{Iachello1987}. It has been shown that triaxiality occurs when one kind of valence nucleons (protons or neutrons) corresponds to particles (described by prolate irreps of SU(3)), while the other one corresponds to holes (described by oblate irreps of SU(3)).  

From Table I and references therein \cite{SDANCA2017Sar} we see that within the $sd$, $pf$, $sdg$, $pfh$, $sdgi$  shells oblate irreps occur within the even-even intervals 8-10, 14-18, 24-28, 34-40, 46-54. These imply that within the spin-orbit shells \cite{Martinou2020} 14-28, 28-50, 50-82, 82-126, 126-184 oblate irreps will occur for 22-24, 42-46, 74-78, 116-122, 172-180 particles, while within the 3D-HO shells 8-20, 20-40, 40-70, 70-112, 112-168 shells oblate irreps will occur for 16-18, 34-38, 64-68, 104-110, 158-166 particles respectively.  These intervals define stripes on the nuclear chart, within which triaxiality should be favored. Regarding the intervals related to the 3D-HO, we expect only the first two of them to make a visible contribution, since the shell gap 20 is strong and the subshell gap 40 is also quite strong \cite{Sorlin2008}, in contrast to 70, 112, 168, where practically no gaps are seen. This topic will be the subject of subsequent work. 

\section{Outlook}

Proxy-SU(3) symmetry has led to the development of the semimicroscopic algebraic quartet model (SAQM) \cite{Cseh2020,Hess2021,Berriel2021,Restrepo2023,Restrepo2024} for the description of heavy nuclei in terms of quartets formed by two protons and two neutrons. For nuclei in which valence protons and neutrons occupy the same shell, the proxy-SU(4) symmetry has been recently developed \cite{Kota2024}. The microscopically derived by proxy-SU(3) hw SU(3) irreps have started  being used within phenomenological algebraic models, as the microscopic version of the Bohr-Mottelson collective model bearing the Sp(12,R) symmetry \cite{Ganev2022,Ganev2024}, and the vector boson model (VBM) \cite{Minkov1997,Minkov2024}. The variable $\gamma$, obtained through proxy-SU(3), has been used for introducing intrinsic triaxial deformation in a microscopically derived IBM-1 Hamiltonian \cite{Vasileiou2024}.

\section*{Acknowledgements}

Support by the Bulgarian National Science Fund (BNSF) under Contract No. KP-06-N48/1  is gratefully acknowledged.

\end{document}